\documentclass{aipproc}
\layoutstyle{6x9}
\usepackage{amsmath, amssymb}
\begin{document}

\title{Holgraphy and BMS field theory}
\author{G. Arcioni}{address={The
Racah Institute of Physics, The Hebrew University, \\ Jerusalem 91904, Israel.}}
\author{C. Dappiaggi\footnote{contributing author}\;}{address={Dipartimento 
di Fisica Nucleare e Teorica,\\
Universit\`{a} degli Studi di Pavia, INFN, Sezione di Pavia, \\
via A. Bassi 6, I-27100 Pavia, Italy} 
}
\begin{abstract}
We study the key ingredients of a candidate holographic correspondence in an asymptotically flat spacetimes; in particular we develop the kinematical and the classical dynamical data of a BMS invariant field theory living at null infinity.
\end{abstract}
\maketitle
Since 't Hooft foundational paper \cite{'tHooft}, the holographic principle has 
played a key role in improving our understanding on the nature of gravitational 
degrees of freedom in a quantum field theory over a curved background. 
Originally the principle was proposed in order to solve the apparent information
paradox of 
black holes by means of a theory living on a lower dimensional hypersurface (usually 
the boundary) with respect to bulk spacetime where all the physical information 
of the manifold is encoded. Moreover, motivated by the Bekenstein entropy 
formula, the density of datas on the ``holographic screen'' should not exceed 
the Planck density; this implies that there is a high redundancy in the way we 
usually count degrees of freedom in a quantum field theory since if we excite 
more than $\frac{A}{4}$ degrees of freedom, we end up with a black hole.

A way to explicitly realize the holographic principle is based on the reconstruction of the bulk starting directly from boundary data 
explaining how they are generated, their dynamics and mainly how they can reproduce classical spacetime geometry. 
The main example is the AdS/CFT correspondence \cite{Aharony} (see \cite{deBoer} for a recent review) where, 
in the low energy limit, a supergravity theory living on $AdS_d\times M^{10-d}$ 
is a SU(N) conformal gauge field theory living on the boundary of AdS. The whole approach is based on the assumption of the equivalence
 of partition sum of gravity and gauge theory once asymptotically AdS boundary 
 conditions are imposed on the bulk spacetime. Thus it seems rather natural to  
investigate whether we could find a similar holographic description once we choose a 
 different class of manifolds and thus of boundary conditions and in particular we
 will refer to asymptotically flat (AF) spacetimes, a scenario where this problem has been
 addressed only in the last few years.

Different approaches have been proposed: in a recent one 
\cite{deBoer2} \cite{Solodukhin}, a Minkowski background is considered and it is 
divided in AdS and dS slices; the idea is to apply separately both AdS/CFT and 
dS/CFT correspondence and then patch together the results. This scenario is 
interesting but it is limited up to now only to the flat case and it is 
unclear how to extend it to a generic asymptotically flat spacetime\footnote{Another 
recent paper \cite{Alvarez} suggests to relate holography in a Ricci flat 
spacetime to a Goursat problem i.e. a characteristic problem in a Lorentzian 
setting. }.

In \cite{Arcioni}, \cite{Arcioni2}, instead, we have explored the 
holographic principle in four dimensional AF spacetimes through the study of the 
asymptotic symmetry group at null infinity $\Im^\pm\sim\mathbb{R}\times S^2$,
namely the Bondi-Metzner-Sachs group (BMS). In Penrose intrinsic 
construction of $\Im^\pm$ and in the so-called Bondi reference frame
$(u=t-r,r,z,\bar{z})$, the BMS is the diffeomorphism group preserving the
degenerate boundary metric $ds^2=0\cdot du^2+d\Omega^2$ i.e.

\begin{center}
$z\to z^\prime=\frac{az+b}{cz+d}\quad ad-bc=1,\qquad
u\to u^\prime=K(z,\bar{z})(u+\alpha(z,\bar{z}))$
\end{center}

where $\alpha$ is an arbitrary square integrable function over $S^2$ and $K$ is a
fixed multiplicative function \cite{Arcioni}. Thus the BMS group has the structure of
the semidirect product 
\begin{equation}\label{bms}
BMS_4=SL(2,\mathbb{C})\ltimes L^2(S^2).
\end{equation}
The underlying strategy we have followed has been both to point
out the differences between the flat and the AdS scenario and to study the key 
ingredients of a field theory living on $\Im^\pm$ invariant under  (\ref{bms}). 
The first step (described in details in \cite{Arcioni}) has been to construct, 
by means of pure group theoretical techniques, the kinematical data of the 
boundary theory i.e. the full particle spectrum of a BMS field theory. Following 
Wigner approach \cite{Group2} and working in a momentum representation, we can 
introduce a \emph{BMS invariant field} as a map $\psi^\lambda:L^2(S^2)\to\mathcal{H}^\lambda$ where $\mathcal{H}^\lambda$ is a suitable target Hilbert space; $\psi^\lambda$ transforms under the action of $g=(\Lambda, p(z,\bar{z}))\in BMS_4$ through 
\begin{equation}\label{covariant}U(g)\psi^\lambda(p^\prime)=e^{i\int\limits_{S^2}\frac{dzd\bar{z}}{1+|z|^2}p(z,\bar{z})p^\prime(z,\bar{z})}D^\lambda(\Lambda)\psi^\lambda(\Lambda^{-1}p^\prime),
\end{equation}
where $D^\lambda(\Lambda)$ is a unitary $SL(2,\mathbb{C})$ representation. At the same time we can introduce the \emph{induced wave function} which transforms under a unitary and irreducible representation of $BMS_4$  constructed from the set of BMS little groups $L=H\ltimes L^2(S^2)$ where $H$ is any compact subgroup of $SL(2,\mathbb{C})$ i.e. 
\begin{equation}\label{induced}
\tilde{\psi}^j:\mathcal{O}\sim\frac{SL(2,\mathbb{C})}{H}\to\mathcal{H}^j,
\end{equation}
where $\mathcal{H}^j$ is a suitable target Hilbert space.  Thus, the particle spectrum of a BMS invariant theory is characterized by fields labelled by each different little group $H$ and by a mass identified through the following projection:
\begin{equation}\label{mass}\pi(p(z,\bar{z}))=\pi\left(\sum\limits_{l=0}^\infty\sum\limits_{m=-l}^lp_{lm}Y_{lm}(z,\bar{z})\right)=(p_{00},...,p_{11})=p_\mu\Longrightarrow m^2=\eta^{\mu\nu}p_{\mu}p_{\nu}.
\end{equation}
The key components of the overall picture can be resumed in the following table:

\begin{center}
\begin{tabular}{|c|c|c|}
\hline
BMS field & labels & orbit $\mathcal{O}$\\
\hline
SU(2) & $m^2>0$ & $\frac{SL(2,\mathbb{C})}{SU(2)}\sim\mathbb{R}^3$\\
\hline
SO(2) & $m^2=0$ & $\frac{SL(2,\mathbb{C})}{SO(2)}\sim\mathbb{R}^3\times S^2$\\
\hline
\end{tabular}
\end{center}
In Wigner's approach, the equations of motion for each BMS particle arise as a 
set of constraints to impose on (\ref{covariant}) in order to reduce its form to 
that of (\ref{induced}). In detail each field has to satisfy an orbit constraint
reducing $L^2(S^2)$ to each $\mathcal{O}(H)$, a mass equation enforcing 
(\ref{mass}), and eventually since $\mathcal{H}^\lambda$ is bigger than 
$\mathcal{H}^j$ an orthoprojection reducing the components in excess in 
$\psi^\lambda$. To give an example, let us consider the BMS $SU(2)$ scalar field 
in the covariant form $\psi:L^2(S^2)\to\mathbb{R}$ and in the induced one 
$\tilde{\psi}:\frac{SL(2,\mathbb{C})}{SU(2)}\to\mathbb{R}$; the equations of 
motion are (BMS Klein-Gordon equation):

\begin{center}
$[p(z,\bar{z})-\pi(p)]\psi(p)=0\;\;\textrm{and}\;\;[\eta^{\mu\nu}\pi(p)_\mu\pi(p)_\nu-m^2]\psi(p)=0.$
\end{center}

\noindent Starting from the above data, we have studied in \cite{Arcioni2} the 
classical dynamic of each BMS field and in 
particular we have addressed the question if the equations of motion could be derived from a variational principle. Working 
in an Hamiltonian framework, we have constructed for each field the covariant phase space i.e. the set of dynamically 
allowed configurations and, through symplectic techniques, the set of possible energy functions. Surprisingly we have found 
a 1:1 correspondence between the covariant phase space either of BMS $SU(2)$ fields and Poincar\'e SU(2) fields 
either of BMS $SO(2)$ fields and Poincar\'e $E(2)$ fields. From an holographic point of view this implies that the boundary 
theory already at a classical level encodes all the information from the bulk (Poincar\'e invariant)
theory; nonetheless this is not sufficient to claim about an holographic correspondence since the latter is fully manifest 
at a quantum level and we still lack a proper understanding of the quantum BMS field theory not to mention 
a way to reconstruct bulk data from boundary ones. On the other hand we also suspect that
the BMS field theory is related to the IR sectors of pure gravity and this
subject is currently under investigation \cite{Arcioni3}.

\end{document}